\documentclass[usenatbib]{mn2e}
\usepackage{color}
\usepackage{epsfig}
\usepackage{amsmath,amssymb}
\newcommand{\hMsol}{{\>h^{-1}\rm M}_\odot}
\newcommand{\hMpc}{{\>h^{-1}\rm  Mpc}} 
\newcommand{\hkpc}{{\>h^{-1}\rm kpc}}
\newcommand{\kpc}{{\>\rm kpc}}
\newcommand{\kms}{{\>\rm km\,s^{-1}}}
\title{Ram pressure statistics for bent tail radio galaxies}
\author[Mguda et al.]
{Zolile Mguda$^{1, 3}$, Andreas Faltenbacher$^2$,Kurt van der Heyden$^{1}$, Stefan Gottl{\"o}ber$^4$,
\and Catherine Cress$^{5,7}$, Petri Vaisanen$^{3}$, Gustavo Yepes$^6$\\\\
{$ ^{1}$Astrophysics, Cosmology and Gravity Centre (ACGC), Department of Astronomy,}\\ 
{\ University of Cape Town, Private Bag X3, Rondebosch 7701, South Africa.}  \\
{$ ^2$Department of Physics, University of Witwatersrand, Braamfontein, 2000} \\
{$ ^3$South African Astronomical Observatory (SAAO), PO Box 9, 7935 Observatory, Cape Town,
South Africa.} \\
{$^4$Leibniz Institute for Astrophysics, An der Sternwarte 16, 14482, Potsdam} \\
{$ ^5$CHPC, CSIR, 15 Lower Hope Rd, Rosebank, 7700, South Africa}\\
{$^6$Departamento de F\'\i sica Te\'orica M-8, Universidad Aut\'onoma de  Madrid, Cantoblanco 28049 Madrid Spain}\\
{$ ^7$Physics Department,  University of the Western Cape, Modderdam Rd, Bellville, 7535, South Africa}\\
}

\begin{document}
\date{\today}
\maketitle
\begin{abstract}
In this paper we use the MareNostrum Universe Simulation, a large scale, 
hydrodynamic, non-radiative simulation in combination with a simple abundance
matching approach to determine the ram pressure statistics for bent radio sources (BRSs).
The abundance matching approach allows us to determine the locations of all galaxies 
with stellar masses $\geq 10^{11}\hMsol$ in the simulation volume.  
Assuming ram pressure exceeding 
a critical value causes bent morphology, we compute the ratio of
all galaxies exceeding the ram pressure limit (RPEX galaxies) relative
to all galaxies in our sample. According to our model 50\% of the RPEX galaxies at  
$z = 0$ are found in clusters with masses larger than
$10^{14.5}\hMsol$ the other half resides in lower mass clusters.
Therefore, the appearance of bent tail morphology alone does not put tight constraints 
on the host cluster mass. In low mass clusters, $M \leq 10^{14}\hMsol$, RPEX 
galaxies are confined to the central $500\hkpc$ whereas in clusters of $\geq 10^{15}\hMsol$ they 
can be found at distances up to $1.5\hMpc$. Only clusters with masses $\geq 10^{15}\hMsol$ 
are likely to host more than one BRS. Both criteria may prove useful in
the search for distant, high mass clusters.
\end{abstract}

\section{Introduction}
\label{sec:intro}
The unprecedented sensitivity and sampling speed of the new generation  
of radio telescopes, such as LOFAR and SKA, will yield an enormous number 
of new low-luminosity nearby as well as luminous distant radio sources. 
A significant fraction of the continuum sources are expected to be radio 
galaxies with double radio lobes formed by Active Galactic Nuclei (AGN), 
which can be detected at very large redshifts. Roughly 1\% of the known 
radio galaxies show bent tail morphology , i.e., bending of the two 
lobes in the same direction \citep[e.g.,][]{Blanton2000, 
    Wing2011}. A gallery of bent tail radio sources (BRSs) can be
found in Figure 1 of \cite{Blanton2001}. The bent morphology is
thought to be a result of the ram  
pressure exerted on the lobes when the ambient gas density is high and
the relative velocity - between the ambient gas and the radio galaxy - is
large. Both conditions are readily fulfilled within
the intra cluster medium (ICM), 
i.e, the high-density, plasma-filled regions of galaxy
clusters. Therefore, bent tailed radio sources may  
provide an efficient detection mechanism for clusters of galaxies 
even at high redshifts. 

The evolution of the number density of high mass clusters is an interesting 
cosmological probe. According to the hierarchical structure formation 
paradigm the most massive structures of the Universe form late. The exact 
timing depends on the cosmological model. A single very massive cluster can 
pose a potential challenge to the currently favoured concordance model if it 
is found at an incompatibly high redshift \citep{Allen2004, Ade2013XX}.
BRSs have been successfully used as tracers of galaxy clusters for
redshifts up to $\sim 1$ \citep[e.g.,][]{Hintzen1978, Deltorn1997, Blanton2001}.

In this paper we {\it assume} that the ram pressure
  due to the motion of galaxies through the ICM, $p_{\rm
    ram}=\rho_{\rm ext} v_{\rm gal}^2$, is the primary cause 
  of bent radio morphology \citep[e.g.,][]{Miley1972,
    Venkatesan1994}. In other words, all BRSs are 
  radio sources exposed to ram pressure above a critical
  value. Buoyancy effects 
  and other interactions which potentially also could cause bent
  morphology are left for a later study. The main argument
  in the paper can be summarised as follows:
  \begin{enumerate}
    \item \label{i} The study focuses on radio galaxies with jet linear size 
    greater than 50kpc which populate the high mass end of the galaxy 
    mass function ($M_{\rm stellar} \gtrsim 10^{11}\hMsol$).  
  \item \label{ii} We assume that all extended radio sources of that size require a
    minimum critical ram pressure in order to show bent morphology. This is a major 
    simplification of the jet physics that is geared towards allowing for a statistical 
    approach to the study of the BRS environments.
  \item \label{iii} The critical ram pressure used to match the {\it
    relative numbers} of bent sources observed is set based on the
    bending equation of extragalactic jets.
  \item \label{iv} The {\it relative numbers} of bent sources expected
    in different environments are analysed based on the MareNostrum
    Universe Simulation.  
\end{enumerate}

  Point ~\ref{i} allows us to apply a simple abundance matching 
  scheme to  determine the locations of potential radio lobe galaxies 
  in the simulation as discussed in \S~\ref{sec:abundance}. 
  We focus only on large radio lobes  
   ($\geq 50\hkpc$): firstly, because lobes are observed to reach megaparsec 
   scales \citep[e.g., 3C236, NGC315, NGC6251
     in][]{3C236,NGC315,NGC6251} which makes them more easily
   detectable at high redshifts; and secondly, to make sure that the
   resolution of the simulation is  
   sufficient to accurately determine the ram pressure at this scale. 
   A side benefit is that we circumvent the discussion of small jet 
   sizes at the beginning of the radio-loud phase as well.

   Point ~\ref{ii} relies on
  a definition of `bent morphology' which, in this work, is based on
  visual inspection of radio sources with bent morphology 
   such as presented in \cite{Blanton2001}. As a consequence of this we do not
   differentiate between Fanaroff-Riley type I and II
   \citep{Fanaroff1974}. Observations show that radio galaxies with
   jet sizes $\geq 40\hkpc$ are evenly distributed between FR I, FR II with FR
   I/II hybrid types being a small fraction of the sample\citep{Best2009}. However, we note that FR I galaxies may be more
   easily bent even at small distances from the core since they have
   broader jets with internal turbulence making them less stiff. In
   contrast, the strongly collimated, highly relativistic FR II jets
   are very difficult to bend. In that case bent tail morphology
   manifests itself in the downstream shift of the radio lobes at the
   end of the jets.

Points ~\ref{iii} and \ref{iv} emphasize that we
   focus on the relative number of BRSs with respect to all radio
   sources of comparable size. In more detail, we use the MareNostrum
   Universe Simulation to determine the environment which is most 
conducive to the formation of BRSs. For that we assume that the
  ratio of BRSs to all extended radio sources (both $\geq 50\hkpc$)
  is equal to the ratio of galaxies which exceed the ram pressure
  criterion ({\it RPEX galaxies}) to all massive galaxies (both $\geq
  10^{11}\hMsol$). Therefore, we investigate the relative abundances
  and locations of RPEX galaxies as a proxy for BRSs. In particular, 
  we focus on the relative abundances as a function of host cluster
  masses down to $10^{13}\hMsol$. The investigation of relative
  numbers also allows us to circumvent the detailed discussion of the
  radio source duty cycles of these source which is an inherently difficult problem.

The paper is structured as follows. In \S\ref{sec:method} we discuss
the methodology. We introduce the simulation, review halo finding and
abundance matching approaches and explain how the ram pressure is
determined. The results are presented in \S\ref{sec:results} followed
by the conclusion in \S\ref{sec:conclusion}.
\section{Methodology}
\label{sec:method}
We use the MareNostrum Universe simulation, which is a non-radiative 
hydrodynamical simulation within a cube of $500\hMpc$ per side to 
determine the ram pressure at the cores of dark matter halos where massive 
elliptical galaxies are anticipated to reside. The ram-pressure is computed 
as the product of the ambient density times the square of the galaxy's relative 
velocity, $p_{\rm ram}=\rho_{\rm ext} v_{\rm gal}^2$. If the
ram-pressure exceeds a critical limit, the radio lobes associated with
the galaxies are assumed to be bent in a downstream direction. In the 
following sections we introduce the simulation, including the halo finding 
procedure, the abundance matching approach which assigns galaxies to the cores 
of the dark matter halos and the determination of the ram pressure
which we assume is the cause for the bent tail morphology.  
\subsection{Simulation and halo finding}
\label{sec:halos}
The present analysis employs the $z = 0$ snapshot of the MareNostrum
Universe simulation \citep{Gottloeber2006} which is based on the
entropy conserving GADGET-2 code \citep{Springel2005} to model both 
dark matter and gas components assuming a concordance cosmology with the
following parameters: $\Omega_m=0.3$, $\Omega_{\Lambda} = 0.7$,
$\Omega_b = 0.045$, $\sigma_8 = 0.9$ and $h=0.7$. These values are slightly 
different to the present values from the Planck survey
\citep{Ade2013XVI} but this should not affect the basic results. The
simulation follows the non-linear evolution of the gas and dark matter density
fields from $z = 40$ to the present within a co-moving box of
$500\hMpc$ on each side. Each component is resolved with
$1024^3$ particles resulting in gas and dark matter particle
masses of $1.5 \times 10^9\hMsol$ and $8.3 \times 10^9\hMsol$,
respectively. Only adiabatic gas dynamics are considered, i.e, 
radiative processes, star formation and feed back processes are not
included. The spatial force resolution was set to an equivalent Plummer 
gravitational softening of $15 h^{-1}$ comoving kiloparsecs. The SPH smoothing 
length was set to the distance to the 40th nearest neighbour of each SPH 
particle.

For the halo and subhalo identification we utilize a hierarchical
friends of friends algorithm \citep{Klypin1999} with
progressively shorter linking lengths of $b_n=b/2^n$ with $n =
0, 1, 2, 3$. Our choice of the basic linking length of $b=0.17$ times the
mean particle separation is commonly used to define particle
concentrations with virial overdensity. Structures are identified 
based on their dark matter particle distribution only. We use $b$, or
equivalently $n=0$, to find the host halos and $n=3$ for
subhalos. These $b_3$ - subhalos (in the following simply referred to
as {\it subhalos}) show overdensity of roughly 512 times
the virial overdensity. We identify their centres of mass as possible
locations of galaxies. In cases where distances between subhalos
(galaxies) and the centre of the host halos are required we 
compute the distance between the location of a subhalos and the centre
of mass of the nearest parent halo, i.e., the halo extracted with
a linking length of 0.17 times the mean particle separation.   
\subsection{Abundance matching}
\label{sec:abundance}
\begin{figure}
  \epsfig{file=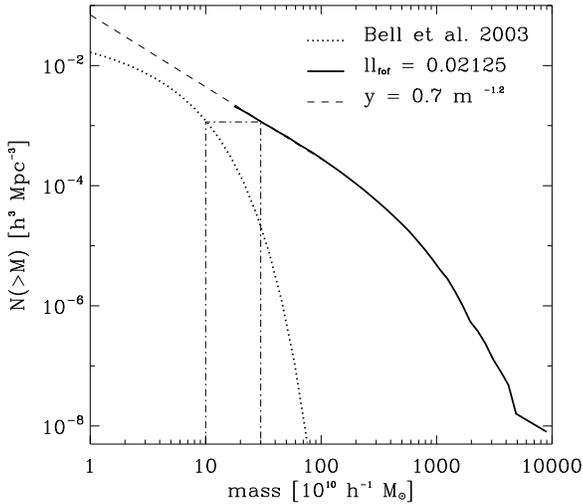,width=0.95\hsize}
  \caption{\label{fig:match}
    The cumulative mass functions based on the ($b_3$) subhalo masses
    (thick solid line) and the stellar masses of SDSS/2MASS galaxies  
    derived by Bell et al. (2003) (thick dotted line). The dashed line
    is an analytical extension of the halo mass function which is
    included as a guide for the eye. The dash-dotted box illustrates the
    abundance matching procedure between subhalos and 
    galaxies. Equal number densities are
    obtained for the SDSS/2MASS galaxies with stellar masses $\geq 
    10^{11}\hMsol$ and subhalos with masses $\geq 3 \times
    10^{11}\hMsol$. Given that the dark matter particles have masses
    of $8.3 \times 10^9\hMsol$ the subhalo mass limit can be
    translated into minimum particle-number limit of 36.   
  }
\end{figure} 
Abundance matching is a technique used to populate theoretically derived dark matter 
halo distributions with observed galaxy properties based on the simultaneous 
gauging of their cumulative mass (or luminosity) functions. We employ a 
simple abundance matching technique similar to the approach described in 
\cite{Conroy2006} to assign stellar masses to the subhalos. For the 
present study we use the galaxy stellar mass function given in
\cite{Bell2003} which employs the {\it ugriz} magnitudes from the Sloan
Digital Sky Survey Early Data Release (SDSS-EDR;
\citealt{Stoughton2002}) in conjunction with K-band fluxes from 
the Two Micron All Sky Survey Extended Source Catalog (2MASS;
\citealt{Jarrett2000}) to translate the galaxy luminosities into
stellar masses.

An illustration of the abundance matching scheme is given
in Fig.~\ref{fig:match}. This figure shows the cumulative dark matter
(bold solid line) and stellar (bold dotted line) mass functions based
on the subhalo and the  SDSS/2MASS galaxy sample, respectively. The
intersections between a horizontal dash-dotted line and the two
distributions indicate the dark matter and stellar masses for which
the cumulative number densities of the two distributions match. 
The dash-dotted box in Fig.~\ref{fig:match} 
indicates that equal number densities are obtained for the SDSS/2MASS
galaxies with stellar masses $\geq 10^{11}\hMsol$ and subhalos
with masses $\geq 3 \times 10^{11}\hMsol$. Galaxies with stellar mass 
$\geq 10^{11}\hMsol$ predominantly show elliptical appearance \citep{Bell2003}. Given a
dark matter particle mass of $8.3 \times 10^9\hMsol$ the subhalo mass limit can be
translated into minimum particle-number limits of 36.

Finally, the connection between galaxy mass and radio-loudness needs to
be determined. For that we make use of the following observationally
established relations. A tight correlation is found between a galaxy's bulge luminosity
and the mass of its Supermassive Black Hole
\citep[SMBH, ][]{Ferrarese2002} suggesting that the most massive galaxies, which
are ellipticals, host the most massive black holes. SMBHs can trigger
AGN activity. \cite{Kauffmann2003} showed that the distributions of
sizes, concentrations and stellar surface mass densities of elliptical
galaxies with and without AGNs are very similar. \cite{Best2007}
found the radio-loud fraction of AGNs to increase from 0.01\% for
galaxies of stellar mass $3\times 10^{10} \rm{M_\odot} $ to greater
than 30\% for galaxies of stellar mass above $5\times 10^{11}
\rm{M_\odot}$. These findings suggest that, besides the intermittent
nature the AGN activity, AGNs are a random subsample of early type
galaxies of which a given fraction (depending on galaxy mass) is
radio-loud. 

Our results are based on the subhalo sample which corresponds
to stellar masses $\geq 10^{11}\hMsol$ where the percentage of
extended radio-loud galaxies is expected to be substantial. In the
following we will refer to these subhalos as {\it galaxies} and
compute the ratio of RPEX galaxies (see \S~\ref{sec:intro}) to all
galaxies.
\subsection{Ram pressure limit}
\label{sec:limit} 
For the jets of radio-loud AGN to be visibly bent, the ram pressure
caused by the jet on the ambient medium must be of the same order of
magnitude as the ram pressure caused by the motion of the galaxy
through the ambient medium. This gives rise to the following bending  
equation \citep{Begelman1979,Jones1979,Burns1980}:
\begin{equation}\label{equ:curve}
\frac{\rho_{\rm jet}\ v_{\rm jet}^2}{R}=\frac{\rho_{\rm ext}\ v_{\rm gal}^2}{h} , 
\end{equation}
where $h$ is the cylindrical radius of the jet and $R$ is the radius
of curvature of the jet. The values $\rho_{\rm jet}$, $\rho_{\rm ext}$, $v_{\rm jet}$ and
$v_{\rm gal}$ denote the density within the jet, the density of the 
ambient medium, the jet speed and the velocity of the galaxy relative
to the ambient medium. 

An alternative approach as outlined in \cite{Morsony2013} is to
rewrite Eq.~\ref{equ:curve} as follows: 
\begin{equation}\label{equ:limit}
P_{\rm ram} = {h \over R}\ P_{\rm jet} = {h \over R}\ \alpha\ P_{\rm min}\ ,  
\end{equation}
where $P_{\rm jet}$ is the pressure within the jet and $P_{\rm ram} =
\rho_{\rm ext} v_{\rm gal}^2$, $R$ and $h$ are the same quantities as
used in Eq.~\ref{equ:curve}. We will use Eq.~\ref{equ:limit} to
determine the ram pressure limit above which we assume that radio
galaxies show bent tail morphology. The internal pressure, $P_{\rm
  jet}$, can be approximated by the minimum synchrotron pressure, 
$P_{\rm min}$, as outlined in \cite{ODea1985}. It gives a lower limit
for the internal pressure due to the presence of entrained material from the
surroundings \citep{Bicknell1984, Laing2002, Croston2008}
therefore we introduce the parameter $\alpha > 1$. Since $h/R < 1$ we
assume that $h\alpha / R \approx 1$. 
This choice is substantiated by the fact that we obtain roughly the
same ratio of BRSs to all radio lobe galaxies as observed
\citep[][ see also \S~\ref{sec:deter}]{Blanton2000} .
For $P_{\rm min}$ we use the
observed values presented in Table~1 of \cite{Freeland2011} based on
which we determine a mean value of $2.4\times10^9\ \rho_b\ \rm km^2
s^{-2}$ ($\rho_b$ being the cosmic baryon density). The minimum and
maximum values are $1.0\times10^9$ to $4.1\times 10^{9}\ \rho_b\ \rm
km^2 s^{-2}$. This range encompasses the minimum ram pressure required
to produce bent tail morphology. We will employ it for the
establishment of the ram pressure statistics for bent tail radio
galaxies. 

\subsection{Determination of Ram Pressure}
\label{sec:deter}
\begin{figure}
  \epsfig{file=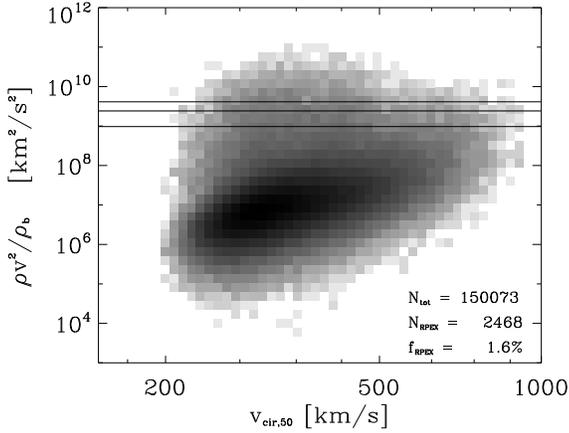,width=0.95\hsize}
  \caption{\label{fig:rps}
    Distribution of ram pressure at the location of some 150000
    subhalos  as a function of their circular velocity based on the total
    matter content within a radius of $50\hkpc$ from the centre of
    mass. Individual data points are gridded and the cell values are
    scaled logarithmically. The horizontal lines indicate the range of
    ram pressure values above which bending of the radio jets is
    expected to occur (see discussion of Eq.~\ref{equ:limit} in the
    text). For instance, according to the ram pressure limit
    indicated by the middle (thick) line 1.6\% of all
    galaxies with stellar massed above $10^{11}\hMsol$ live under
    conditions where ram pressure is sufficiently large to cause bent
    tail morphology (if the galaxies happen to form radio jets during
    this time).}  
\end{figure}
Equation~\ref{equ:limit} requires the determination of the density of the
ambient ICM and the velocity of the galaxy relative to it from the simulation. 
We compute these quantities in the following way: First we determine
the centre of mass and the bulk velocity of the subhalo based
on its dark matter component only; then we compute the average ICM density
and relative velocity within a halo-centric shell with lower and upper
bounds of $50\hkpc$ and $100\hkpc$, respectively. The lower bound
is chosen to be comparable to the smallest radio lobes we consider. 

Empirical relations show that galaxies with stellar masses of $\geq 10^{11}\hMsol$
host black holes with $M_{\rm BH } \geq 10^8 M_{\odot}$ \citep{McLure2002}. Black 
holes of that mass produce jets which propagate out to $\gtrsim 50 \kpc$. 
\cite{Best2009} shows that 70\% of the extended radio-loud sources in the 
nearby Universe are larger than $50\hkpc$. Provided that the ratio between 
radio-loud galaxies with linear jet size larger than $50\hkpc$ and
galaxies more massive than $10^{11}\hMsol$ is independent of
environment we then set out to determine the fraction of BRSs among
all radio-loud galaxies with lobes larger than $50\hkpc$. The upper
bound, $100\hkpc$, is chosen to allow for a sufficiently large volume
to determine the ambient ICM density while minimizing non local
contributions. 

Large relative velocities indicate that the gaseous halo of the galaxy has
been stripped at least outside of $50\hkpc$. In this case, it gives the
velocity with which the ambient ICM is streaming past the galaxy. In
situations where stripping is not efficient, the gas density within 50
to 100 $\hkpc$ may be dominated by the galaxy's own gaseous halo and
we do not measure the {\it ambient} ICM density. Since, in this case,
the relative velocity between the dark matter core and gas in the shell is
expected to be rather small, the resulting ram pressure is small too
and our approach does not predict bent tail morphology. 

Figure ~\ref{fig:rps} displays the ram pressure for the subhalos 
as a function of circular velocity based on the total mass within a
halo-centric sphere of $50\hkpc$. The ram pressure limits as discussed
in the context of Eq.~\ref{equ:limit} are indicated by the horizontal
lines. The application of the $2.4\times10^9\ \rho_b\ \rm km^2 s^{-2}$
limit indicates that for 2468 out of $1.5 \times 10^5$ subhalos
(1.6\%) the ram pressure is sufficient to bend the tails of radio
galaxies if the galaxies undergo a radio-loud phase during this
time. In the following we refer to these objects as {\it RPEX
  galaxies}. For the lower and upper bounds shown in
Figure.~\ref{fig:rps} we obtain 1.0\% and 3.2\%, respectively.

As discussed above, the fraction of RPEX
galaxies to the total number of galaxies ($m_{\rm star} \geq
10^{11}\hMsol$) approximates the fraction of bent tail sources among
all radio-loud galaxies with lobes larger than $50\hkpc$. Accordingly,
1.6\% of all radio-loud galaxies with radio lobes larger than
$50\hkpc$ show bent tail morphology due to the ram pressure exerted by
their environment. This compares very favourably with observational
results of the $3000 \rm{degree^2}$ FIRST survey, which found 384
BRSs out of a total sample of 32000 double and multiple radio sources
\citep{BlantonPhDT} giving a fraction of 1.2\%. 

One concern regarding the comparison with observations may be the presence
of projection effects. Two different alignment effects may have a bearing 
on the predictions from the simulations: firstly, the alignment between 
jet axis and the line of sight and secondly, the alignment between
the galaxy's velocity and the line of sight. The first case does not
affect our predictions because we are interested in the relative
number of bent tail radio galaxies with respect to all radio lobe
galaxies larger than $50\hkpc$ in size. A jet orientation parallel to
the line of sight affects bent and non-bent radio galaxies in the same
way and consequently does not change the ratio between the two sets.
In the second case, the bent tails of a radio-lobe galaxies moving
parallel to the line of sight may be observed as straight. If we
assume this case for all galaxies moving with an angle of
$\lesssim 30^\circ$ relative to the line of sight then the simulation
overpredicts the fraction of bent tail radio sources relative to all
radio lobe galaxies by 14\% ($1 - \cos \theta$). Given the inherent
uncertainties in our approach we do not explicitly include this
correction into our results. \cite{Morsony2013} states that orientation 
effects cause the jet lengths to be underestimated by and average of $30-50$ percent.

\begin{figure}
  \epsfig{file=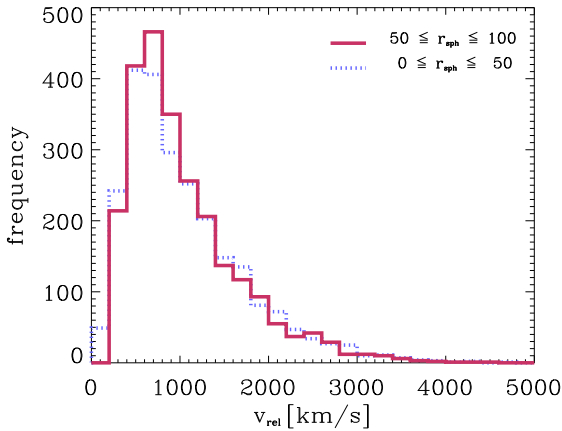,width=0.95\hsize}
 \caption{\label{fig:rpshist}
   Distribution of the relative velocities between the bulk velocity
   of the subhalos and the ambient ICM for all halos which
   fulfill the ram pressure criterion ($P_{\rm ram} \geq
   2.4\times10^9\ \rho_b\ \rm km^2 s^{-2}$). The solid, red line is
   based on the measurement of the velocity within a halo-centric
   shell between $50$ and $100\hkpc$. The dotted, blue
   line gives the relative velocities between subhalos and the ICM
   within a halo-centric sphere of $50\hkpc$. It is shown here to
   provide a consistency check rather than presenting a separate
   result. The similarity between the two graphs indicates that in
   alsmos all cases where the ram pressure criterion is fulfilled the
   ambient ICM streams freely through the subhalo.} 
\end{figure}

The distribution of the relative velocities between the subhalos and the
ambient ICM for all halos which fulfill the ram pressure criterion 
($P_{\rm ram} \geq 2.4\times10^9\ \rho_b\ \rm km^2 s^{-2}$) are
shown in Figure ~\ref{fig:rpshist}. The solid, red line is based on
the measurement of the subhalo velocity relative to the halo-centric
shell between $50$ and $100\hkpc$ which we use to determine the ram
pressure. The relative velocities span a range from $300\kms$ up 
to over $3000\kms$ with a marked peak slightly below $1000\kms$.
The dotted, blue line gives the relative velocities between the
subhalo and the ICM within a halo-centric sphere of $50\hkpc$. The
similarity of the two velocity distributions suggests that most of
the subhalos which fulfill the ram pressure criterion have been 
entirely stripped of their own gas. Consequently, the ambient ICM
streams freely through them, which should also be valid for the
galaxies these subhalos are predicted to host.

We find some galaxies which have fairly low relative velocities with
respect to the ambient ICM ($\lesssim 700 
\rm{kms}^{-1}$) and yet fulfill the bending criterion. This is in
agreement with observations which, in rare cases, report low velocity
environments of BRSs \citep{ODea1985,Jetha2006}. A more detailed
discussion of this observation follows below.  

\begin{figure}
  \epsfig{file=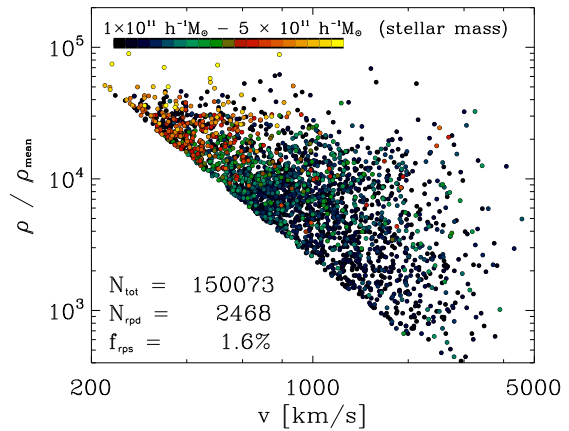,width=0.95\hsize}
  \epsfig{file=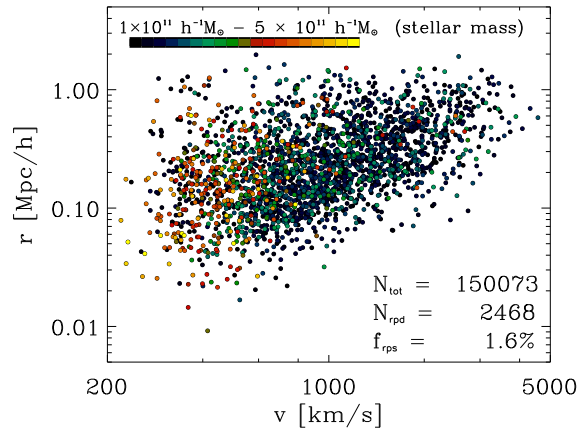,width=0.95\hsize}
  \caption{\label{fig:rhovel}
    {\it Upper panel:} Density versus velocity for subhalos
    which fulfill the bending criterion (Eq.~\ref{equ:limit}). The color coding
    reflects the stellar masses as derived from the abundance matching
    procedure discussed in
    section ~\ref{sec:abundance}. The high
    mass galaxies with low velocities may correspond to wide angle tail  
     sources, whereas the high velocity subhalos are more naturally
    identified as narrow angle tail radio galaxies.
    {\it Lower panel:} Distance to the host halo's 
    centre of mass versus velocity for the same set of subhalos as
    displayed above. There is a trend that high mass 
    galaxies with low relative velocities are located close to the
    centre of the clusters. This is where WATs are observed.}    
\end{figure}
\section{Results}
\label{sec:results}
In order to use BRSs as tracers of clusters we need to be able to
answer the following questions: What is the typical mass of the host
clusters? What are the typical distances between BRSs  and the host
cluster centres? And, how many BRSs are likely to be found within a
single host cluster of a given mass? The subsequent analysis aims to
extract answers to these questions from the MareNostrum simulation.
As emphasized before, we aim to determine the fraction of bent tail radio
galaxies relative to all radio-loud galaxies with lobes larger than
$50\hkpc$. This approach avoids detailed modelling of the jet physics
and allows predictions based on the fraction of RPEX galaxies
relative to all galaxies ($m_{\rm star}\geq 10^{11}\hMsol$).   
\subsection{Slow and fast moving galaxies}
For the following discussion the differentiation between 
narrow angle tail galaxies (NATs) and wide-angle tail galaxies (WATs)
is instructive, therefore a short definition is presented here. NATs, 
which are U-shaped radio sources 
where the angle subtended at the core by the jet arms is less than 
$45^\circ$, have long been believed to be a product  of the
ram pressure experienced by the galaxy moving through the
Intra-Cluster Medium (ICM) at high velocity \citep{Miley1972, ODea1985}. 
This mechanism is inadequate to address the morphology of slower-moving, 
C-shaped WATs without invoking much denser environments. 
WATs have subtended angles that are greater than $90^\circ$ and are 
usually associated with the brightest cluster galaxies (BCGs) that are 
located close to the cores of massive galaxy clusters with only moderate 
velocities relative to the ambient ICM \citep{O'Donoghue1993}. 

As a first step to answering the questions raised above we
plot the RPEX galaxies in the $\rho - v$ plane,
shown in the upper panel of Figure~\ref{fig:rhovel}, 
where $\rho$ is the density of the ambient ICM and $v$ is the relative 
velocity between the galaxy and the ICM. The colour coding in this
figure reflects the stellar mass as indicated by the color bar in the
upper right corner. Yellow points correspond to the most
massive galaxies with stellar masses of $\sim5\times10^{11}$. They are
found predominantly in high density environments with low relative
velocities which can be readily identified as BCGs at the centres of
massive clusters. The black, blue and green dots, which correspond 
to lower mass galaxies display a wider distribution of velocities and
ambient densities. The figure shows that more massive galaxies which 
predominantly reside in high density regions require less extreme 
velocities with respect to the ambient ICM to generate RPEX
galaxies. The apparent line, below which there are no qualifying RPEX
galaxies corresponds to $P_{ram}=2.4 \times 10^9\ \rho_b\ \rm km^2
s^{-2}$.

The lower panel of Figure~\ref{fig:rhovel} displays the distances with
respect to the host cluster's centre of mass versus the velocity
relative to the ambient medium. This figure reveals a clear
correlation between cluster centric distance and the relative velocity
between the RPEX galaxies and the ambient medium. In particular the
most massive galaxies which generally show low velocities relative to
the ambient ICM preferentially inhabit the central regions of the
clusters. That is where observations preferentially find the WATs 
\citep{Burns1996}. Notwithstanding the above correlation, 
there is also a good distribution of RPEX galaxies throughout the 
different cluster-centric distances.

The radio lobes of very massive galaxies which resemble BCGs
are bent by the fact that the galaxies move through the clusters' central,
high density regions with relatively low velocities. Therefore, the
associated radio lobes may fall into the WAT category.
The RPEX galaxies found at the outskirts of the galaxy clusters show
high relative velocities with respect to the ambient ICM and are, therefore, 
expected to show NAT morphology which, again is
  confirmed by observations \citep{ODea1985}. The continuous transition 
from NATs to WATs in Figure~\ref{fig:rhovel} shows that all RPEX galaxies 
belong to the same family and we will not differentiate between them in the 
remainder of the analysis. That does not necessarily mean that the NATs and 
WATs originate from the same physical processes. For instance, it is 
conceivable that the jet physics are different as a result of the different 
environments.

This result illustrates the importance of both the velocity at which
the galaxies are travelling through the ICM and the density of the ICM.
The cluster centres, where the ICM densities are high, are dominated by
BCGs which only move slowly relative to the ICM. \cite{Croft2007} found 
19.7\% of BCGs' in the FIRST Survey to be radio-loud. This
radio-loudness is primarily correlated with the stellar mass of the BCG. 
At larger cluster-centric distances the density profiles of the ICM are the main
determinants of the BRS fraction since the density increases from the
outer radii to the cluster centre by a few orders of magnitude whereas
the average velocities of the galaxies orbiting in the cluster
potential well  only rise by a factor of two or so
\citep[e.g.,][]{Faltenbacher2005}.   
\subsection{Dependence on cluster mass}
\begin{figure}
\epsfig{file=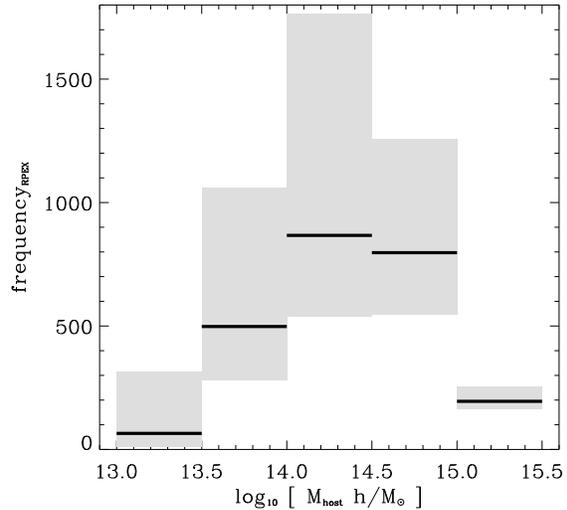,width=0.95\hsize}
 \caption{\label{fig:abs}
   The total number of RPEX galaxies as a function of the host
   cluster's dark matter halo mass. The black lines indicate the 
   results based on the averaged ram pressure limit (middle line in
   Fig.~\ref{fig:rps}) and the shaded areas give the results based on
   the upper and lower limits in Fig.~\ref{fig:rps}.   
   The high mass end of the histogram is a consequence of the finite 
   maximum cluster mass in the simulation volume. The rapid drop at
   the low mass end is given by the fact that the ram pressure is not
   sufficient to bend radio jets in such low mass clusters. The solid, red
   line displays the results for subhalos with $\geq36$ dark matter
   particles which according to the abundance matching procedure 
   (see \S~\ref{sec:abundance}) correspond to galaxies with stellar masses $\geq
   10^{11}\hMsol$.} 
\end{figure}
\begin{figure}
\epsfig{file=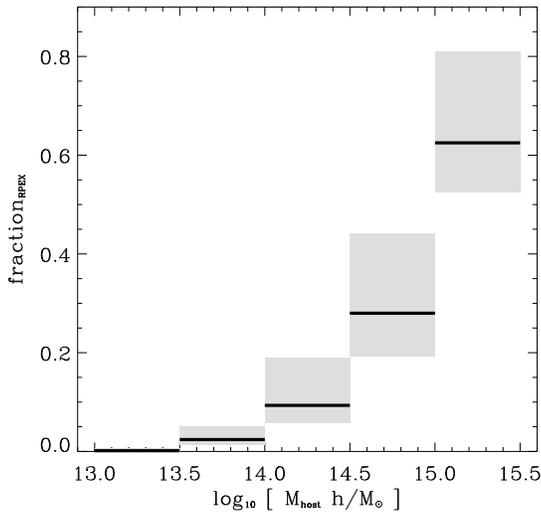,width=0.95\hsize}
 \caption{\label{fig:rel}
   The ratio between RPEX galaxies and the total number of subhalos
   as a function of the host cluster mass. Line styles are 
   the same as in fig.~\ref{fig:abs}.}   
\end{figure}
The mass of the parent cluster affects the jet bending probability in
two ways. Firstly, the velocity dispersion of galaxies is correlated
with the mass of the cluster, i.e.,  galaxies move faster in high mass 
clusters. Secondly, the density of the ICM is correlated with the
cluster mass, i.e., galaxies in high mass clusters experience denser
ICM. Therefore, we expect RPEX galaxies to populate predominantly high mass 
clusters. We quantify this by measuring the number of RPEX galaxies 
with respect to the total number of galaxies as a function of cluster
mass. 

Figure \ref{fig:abs} shows the total numbers of RPEX galaxies as a
function of host cluster mass. We find no RPEX galaxies in clusters
below $10^{13}\hMsol$ (not shown) and only very few in clusters less
massive than $10^{13.5}\hMsol$ although the total number of such low
mass clusters in the simulation box exceeds by far those of higher 
mass. As reported by \cite{Gottloeber2006}, the simulation has more
than 50000 clusters in the $10^{13}-10^{14}\hMsol$ mass range and more
than 4000 clusters with masses greater than $10^{14}\hMsol$. 
Clusters above $10^{14}\hMsol$ provide suitable conditions for
radio jet bending. The absolute numbers of RPEX galaxies peak for 
clusters between $10^{14}$ and $10^{15}\hMsol$. This can be observed
independent of the exact ram pressure limit applied, i.e. the results
of the averaged ram pressure criterion as well as its upper and lower
limits yield similar results. The preponderance of 
low mass clusters and the low number of RPEX galaxies found in them,
points to the inability of low mass clusters to cause the bent
morphology at least for host galaxies with stellar masses above
$10^{11}\hMsol$ as investigated here. 

Figure~\ref{fig:rel} displays the fraction of RPEX galaxies relative to
the total number of galaxies as a function of cluster mass. As
discussed above, this quantity is equivalent to the fraction of bent
tail sources relative to all radio-loud galaxies with lobes larger
than $50\hkpc$. Consequently, a strong correlation 
between cluster mass and RPEX galaxy fraction can be derived. A negligible
fraction of RPEX galaxies is found for host masses below
$10^{13.5}\hMsol$, it rises for clusters with masses between
$10^{13.5}$ and $10^{14.0}\hMsol$ to roughly 5\%. This is a relatively
small fraction. However, it is important to keep in mind that the total
number of RPEX galaxies as shown in Fig.~\ref{fig:abs} is  
relatively large due to the large number of clusters in this mass bin. 
For cluster masses between $10^{15}$ and $10^{15.5}\hMsol$ the
fraction of BRSs rises to $\sim60\%$. Therefore, if such clusters
host more than two radio lobe galaxies half of them are expected to
show bent tail morphology.
\subsection{Distances of BRSs from cluster centres}
\begin{figure}
\epsfig{file=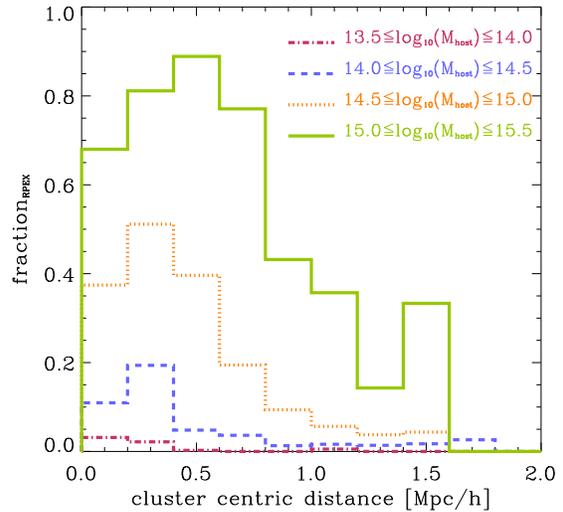,width=0.95\hsize} 
\caption{\label{fig:dist}
  Fractions of RPEX galaxies with respect to total number of
  galaxies as a function of cluster-centric distance for different
  host mass bins as indicated by the labels. More massive clusters
  show a more extended distribution of RPEX galaxies.}  
\end{figure}
To locate galaxy clusters with the help of BRS it is not only
important to predict host cluster masses but also the spatial
distribution of the BRSs with respect to the cluster centre. In other 
words, if a BRS is detected what is the maximum expected distance to
the cluster centre? 

Figure \ref{fig:dist} shows the fractions
of RPEX galaxies with respect to the total number of galaxies as a
function of cluster centric distance and host halo mass. For host
halos above $10^{14}\hMsol$ a slight reduction of the central
($\lesssim 200\hkpc$) fraction of RPEX galaxies can be observed. For those
clusters the fraction of RPEX galaxies peaks between $200$ and $600\hkpc$
with a slight shift of the peak towards larger cluster centric
distances for higher mass host clusters. For host clusters $\ge
10^{15}\hMsol$ 80\% of all double lobe galaxies at a cluster centric
distance of $500\hkpc$ are expected to show bent tail morphology. For 
distances above $\sim600\hkpc$ the fraction of RPEX galaxies drops quickly. 
Our model predicts, that for the most massive clusters 20\% of
all radio-loud galaxies with lobes larger than $50\hkpc$ and stellar
masses $\ge 10^{11}\hMsol$ show bent tail morphology at cluster
centric distances of $\sim 1.5\hMpc$.  

The consequence, for observational strategies to detect clusters by
locating BRSs, is that the cluster centre may be at a distance of
$1.5\hMpc$ from the actual position of the BRS. 
\subsection{Multiple BRS candidates in high mass clusters}
\label{sec:multiple}
\begin{figure}
\epsfig{file=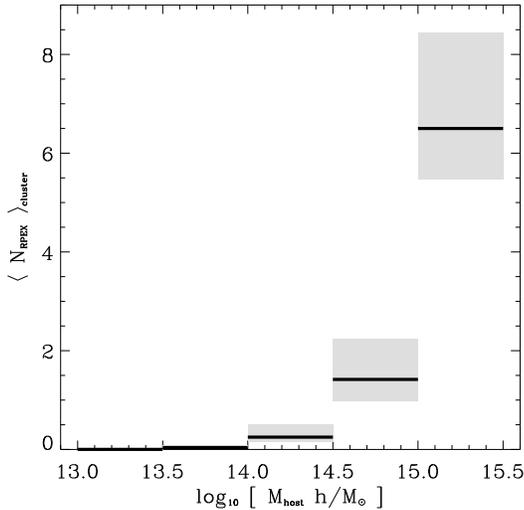,width=0.95\hsize} 
\caption{\label{fig:fc} Ratio of RPEX galaxies within
  clusters of given mass and the total number of clusters in that mass
  bin. One expects to find about 7 RPEX galaxies for clusters
  with masses $\ge 10^{15}\hMsol$. The actual number of
  observable BRSs at a given instant of time depends on the duty
  cycle of the AGN, projection effects and some other biases. 
  Best et al. (2007) found the number of radio-loud
  AGNs to increase to about 30\%  for galaxies with stellar masses
  greater than $10^{11} M_{\odot}$. Therefore, clusters with masses
  above $10^{15}\hMsol$ are likely to host more than one BRSs.}
\end{figure}
\begin{figure}
\epsfig{file=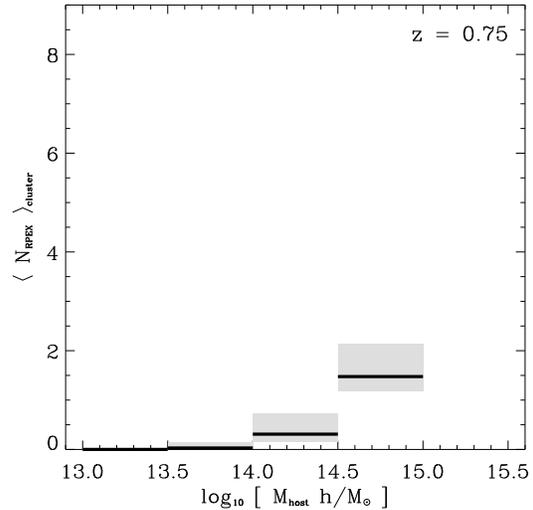,width=0.95\hsize}
 \caption{\label{fig:fc075}
   Same as fig.~\ref{fig:fc} but for a redshift of $z=0.75$. Note that
   that a particle number limit $\geq 36$ for the subhalos may
   correspond to different stellar mass limits due to the higher
   redshift. Nevertheless, qualitatively similar results are
   observed. About two RPEX galaxies are expected to be hosted by
   clusters with masses $\ge 10^{15}\hMsol$.}  
\end{figure}
Figure \ref{fig:fc} shows the average number of RPEX galaxies per
cluster as a function of cluster mass. We find that, while there is
less than one RPEX galaxy in clusters with masses  $\le
10^{14.5}\hMsol$ and about one in clusters with masses between
$10^{14.5}$ and $10^{15}\hMsol$, there are about 7 RPEX galaxies 
expected to be hosted by clusters $\ge 10^{15}\hMsol$. It
depends on the duty cycle of the AGNs how many BRSs actually can be 
observed. \cite{Best2007} found the number of radio-loud AGNs to
increase to about 30\%  for galaxies with stellar masses greater than
$10^{11} M_{\odot}$. Therefore, we can expect that clusters with
masses above $10^{15}\hMsol$ are likely to host more than one
BRSs. This is confirmed by several observations of multiple BRSs in
clusters of galaxies e.g.,Abell 2382, Abell 2538, 
etc. \citep{Ledlow1997}. The appearance of multiple BRSs in close
vicinity may be a used as tracer of high mass galaxy clusters.   

The appearance of multiple BRSs in close
proximity may also be useful for the detection of high
redshift clusters. Figure ~\ref{fig:fc075} gives a coarse
attempt to extend the analysis towards higher redshift. The figure is
identical to Fig.~\ref{fig:fc} except that it is based on the $z =
0.75$ snapshot of the simulation. For an accurate analysis one would
need to redo the abundance matching (\S~\ref{sec:abundance}) with an
observed, $z = 0.75$, stellar mass function. However, we do not do
this in the current paper. Instead, we simply used all subhalos
with more than 36 particles as for the $z = 0$ analysis.   
In Fig.~\ref{fig:fc075}  multiple RPEX galaxies ($N_{\rm
  bent}/N_{\rm cluster} > 1$) can be observed for host halos $\ge
10^{14.5}\hMsol$. Above $10^{15}\hMsol$. we only find one cluster
which is not shown on the plot due to lack of statistics. This is a
preliminary result and a more detailed study based on a higher
resolution simulation will be presented elsewhere.
\section{Conclusions}
\label{sec:conclusion}
Based on the MareNostrum Universe Simulation, we modeled the ram
pressure strength at the expected locations of massive galaxies.  
We applied a simple bending criterion (Eq.~\ref{equ:limit}) which
allowed us to determine whether the radio-lobes associated with these
galaxies will be bent.
If the galaxies exceed the ram pressure limit they are referred to as
RPEX galaxies which indicates that 
they would display bent radio lobes during the radio-loud phase of their 
duty cycle. We derive ratios of
RPEX galaxies relative to all galaxies within the same stellar mass
range ($\geq 10^{11}\hMsol$). Assuming that the AGN duty cycle is independent of environment,
these ratios are equal to the ratios of observed radio
lobes with bent tail morphology to all, straight and bent, radio lobe
sources. For technical reasons, the analysis is restricted to radio lobes larger
than $50\hkpc$.\\ 

The analysis of the velocities and ambient densities of RPEX galaxies
reveals that the most massive galaxies (possibly BCGs), 
which reside at the cluster centres, require only moderate velocities
($\sim 500\kms$) relative to the ambient ICM to obey the bending criterion - 
the ICM densities are high and thus the critical ram pressure is
easily reached without very high relative velocities. This type of
galaxy is expected to host WATs. RPEX galaxies
with lower stellar mass content show a wider distribution of 
relative velocities and ambient densities. Those RPEX galaxies with
high relative velocities are the natural locations of NATs. 
The simulation data show that there is a
continuous transition between WATs and NATs. 
The underlying physical mechanisms are essentially the same.\\ 

Clusters with masses less than $10^{13}\hMsol$ do not host any RPEX
galaxies. We find approximately equal numbers of RPEX galaxies in
clusters less/more massive than $10^{14.5}\hMsol$. Individual low
mass clusters only rarely provide suitable conditions for the 
appearance of bent tail morphology. However, lower mass clusters are much 
more prevalent. Consequently, their contribution to the total budget of 
RPEX galaxies is substantial. Therefore, the appearance of BRSs does not uniquely 
point towards a high mass host cluster.\\  

The RPEX galaxy distribution of cluster-centric distances shows a
marked difference between high and low  mass clusters. The RPEX
galaxies in high mass clusters are found at much larger
distances. This is a consequence of the fact that the ICM of such
systems is much more extended and the velocity dispersion of the galaxies is
higher. Our model predicts that BRSs in high mass clusters can be  
observed as far as $1.5\hMpc$ from the cluster centre. 
For low mass clusters, $\leq 10^{14.5}\hMsol$, the BRS will
preferentially be found within a cluster-centric sphere of $400\hkpc$ 
radius.\\

Finally the appearance of more than one BRS in close proximity
($\lesssim 2\hMpc$) would be a strong indication of high mass host
systems. To what extent this clustered appearance of BRSs may be
useful to detect massive clusters at high redshift needs to be
investigated with higher resolution simulations which provide 
a better mass resolution and thus more accurate results at early
times.\\

The purpose of this work is to lay the ground for investigations of
the statistics of BRSs. The present analysis is incomplete and surely 
requires refinements in many respects. However, the incoming 
large data sets require the development of these statistical tools in
the foreseeable future.

\section*{Acknowledgments}
We thank the anonymous referee for the appreciative reading of our article as well as the 
suggestions that have helped to enhance the quality of our paper.
The Marenostrum Universe simulation was performed at the Barcelona Supercomputing 
Center (Spain). The FOF programs have been developed by Victor Turchaninov during 
a visit at AIP funded by DFG (GO 563/22-1).

ZM thanks the South African Astronomical Observatory (SAAO) and the National 
Research Foundation (NRF) of South Africa.

GY thanks the Spanish\rq{}s MINECO and MICINN for supporting his research through 
different  projects:  AYA2009-13875-C03-02, FPA2009-08958, AYA2012-31101, 
FPA2012-34694 and Consolider Ingenio SyeC CSD2007-0050. He also  acknowledge support 
from the Comunidad de Madrid through the ASTROMADRID PRICIT project (S2009/ESP-1496).
%

%
\end{document}